\documentstyle[prl,aps,epsfig]{revtex}
\def\sqr#1#2{{\vcenter{\hrule height.#2pt\hbox{\vrule width.#2pt
height#1pt \kern#1pt \vrule width.#2pt}\hrule height.#2pt}}}

\newcommand{\be}{\begin{equation}}
\newcommand{\ee}{\end{equation}}
\newcommand{\ben}{\begin{eqnarray}}
\newcommand{\een}{\end{eqnarray}}
\newcommand{\bc}{\begin{center}}
\newcommand{\ec}{\end{center}}

\begin{document}
\draft \twocolumn[\hsize\textwidth\columnwidth\hsize\csname
@twocolumnfalse\endcsname
\widetext

\title{\bf \v{C}erenkov radiation and scalar stars}

\author{S. Capozziello$^{a,b}$\footnote{e--mail:capozziello@sa.infn.it},
G. Lambiase$^{a,b}$\footnote{e--mail:lambiase@sa.infn.it}, and
Diego F.
Torres$^{d}$\footnote{e--mail:dtorres@venus.fisica.unlp.edu.ar}}

\address{$^a$Dipartimento di Scienze Fisiche E.R.
Caianiello, Universit\`a di Salerno, 84081 Baronissi (SA), Italy\\
 $^b$Istituto Nazionale di Fisica Nucleare, Sez. Napoli, Italy\\
 $^c$Instituto Argentino de Radioastronom\'{\i}a, C.C.5,
1894 Villa Elisa, Buenos Aires, Argentina}


\maketitle

\begin{abstract}

We explore the possibility that a charged particle moving in the
gravitational field generated by a scalar star could radiate
energy via a recently proposed gravitational \v{C}erenkov
mechanism. We numerically prove that this is not possible for
stable boson stars. We also show that soliton stars could have
\v{C}erenkov radiation for particular values of the boson mass,
although diluteness of the star grows and actual observational
possibility decreases for the more usually discussed boson masses. 
These conclusions diminish, although do not completely rule out, the
observational possibility of actually
detecting scalar stars using this mechanism, and lead us to consider
other forms, like gravitational lensing.

\end{abstract}

\bigskip

\pacs{PACS no.: 04.40.Dg, 04.40.-b}

\vskip2pc]

\narrowtext

\newpage

\section{Introduction}

One of the first attempts to incorporate a scalar field in the
theory of gravity was done by Brans and Dicke \cite{brans}, who
extended Einstein's theory in order to accommodate Mach's
principle and Dirac's large number hypothesis. In early universe
scenarios, a scalar field (inflaton) could drive inflation through
its potential energy \cite{linde}, and it could give rise to the
formation of topological defects when some fundamental symmetry is
broken \cite{vilenkin}. Furthermore, the presence of a scalar
field (dilaton) naturally appears in the limit of low energy (with
respect to the Planck one) of superstring theory, and it is
responsible for the so-called super-inflation \cite{gasperini}.

All these examples, though in different contexts, strongly suggest
the need to introduce a scalar field in cosmological models. Owing
to these facts, one could investigate if such scalar fields may be
the seed of astrophysical structures or for observable phenomena
that could signal their existence. Objects made up of scalar
massive particles were introduced by Kaup \cite{kaup} and Ruffini
and Bonazzola \cite{ruffini} as a new type of star model, the
so-called boson star. Such objects are macroscopic quantum states
that are gravitationally prevented from collapsing (in opposition
to black holes) by the Heisenberg uncertainty principle, it
keeps scalar particles from being localized to within their
Compton wavelength.

In a recent work \cite{LS}, Liddle and Schunck studied the
gravitational redshift of the radiation emitted within a boson
star potential and the rotational curves of accreted particles to
assess the possible detectability of these scalar stars. They
conclude that very massive boson stars could look very similar to
an active galactic nucleus with a black hole at its center.
Suggestions in that sense were also given before, by Tkachev
\cite{TKA}. This
is indeed supported by the fact that the solution far away from
the boson star center mimics the Schwarzschild space-time. Later,
moreover, Schunck and Torres \cite{ST} explored to what extent
these properties are exclusive of the boson star potential used in
Ref. \cite{LS}, finding that in fact they are even common to more
generic Lagrangian densities.

In our own galactic center, while the existence of a single large
mass has been favored as the upper bound on its size tighten and
stability criteria rules out complex clusters (see for instance
Ref. \cite{gen}), it is not established that this central mass has
to be a black hole. If Sgr A$^*$ (the super-massive compact object
name) is a black hole, its luminosity should be three order of
magnitude bigger than what actually is. This discrepancy is called
``the blackness problem'', and led to the concept of a black hole on
starvation. However, observational data comes from regions
$4 \times 10^{4}$ Schwarzschild radius away from a black hole of
mass $2.6 \times 10^{6} M_\odot$  -the inferred mass of the
central object-: proofs of the existence of a super-massive black
hole in the center of the galaxy are not conclusive by now. We have
constructed a model of a boson star galactic center and studied some of
its properties elsewhere \cite{BOS-GLX}.

With the aim to search for some new physical effect able to reveal the
presence of scalar stars in the universe, other than the
already considered gravitational
lensing \cite{dabrowski}, we analyze the
possibility that a charged particle, propagating in the
gravitational field generated by such a compact object, could emit
radiation via \v{C}erenkov process.

\subsection{\v{C}erenkov gravitational radiation}

\v{C}erenkov radiation occurs when a fast particle moves through a
medium at a constant velocity $v$, which is greater than the
velocity of light in that medium. Because of the superluminal
motion of the particle, a shock wave is created and this yields to
a loss of energy. The wavefront of the radiation propagates at a
fixed angle \be \cos \theta = \frac
{v_{phase}}{v}=\frac{c/n(\nu)}{v}\ee where $\nu$ is the photon
frequency and $n$ is the refractive index. Only in this direction
do the wavefronts add up coherently. The value $\cos \theta = 1$
corresponds to the threshold for emission. It is clear that $\cos
\theta < 1$ can not be satisfied (and then will be no radiation)
if $n < 1$ (because $v$ always is less than  $c$). See, for
example, Refs. \cite{ginzburg,LONGAIR}.

As is well known, charged particles with acceleration $a$ emit
electromagnetic radiation according to the Larmour's relation
$dE/dt \sim a^2$. However, acceleration due only to a
gravitational field does not produce any electromagnetic
radiation, in agreement with the equivalence principle
\cite{pauli}. This follows because the acceleration induced by
gravity disappears in the local inertial frame, so that the
particle moves in that frame with constant velocity, without
radiating.\footnote{In fact, this is not exactly correct. As
Boulware has shown in Ref. \cite{BOUL} the particle does radiate
but all radiation goes in a region which is in-accessible to the
comoving observer. The discussion whether a uniformly accelerated
particle can radiate or not, and issues concerning the violation
of the equivalence principle for charged particles are being
discussed since a long time. See, for instance Refs.
\cite{gasperini2,PARROTT}.} Nevertheless, a particle moving with
constant velocity through an external medium (even a gravitational
field) can emit \v{C}erenkov radiation. This possibility, i.e.
that an external gravitational field acts as an effective
refractive index for light, has been recently analyzed by Gupta,
Mohanty and Samal \cite{gupta}. They showed that the background
gravitational field has an effective refractive index given by
\begin{equation}\label{1.1}
  n^2_{\gamma}(k_0)=|\eta^{00}|\left(1-\frac{R^i_{\phantom{i}i}}{|\eta^{00}|
  k_0^2}\right)\,,
\end{equation}
where $R^i_{\phantom{i}i}$ is understood as the sum on the spatial
indices of the Ricci tensor $R^{\mu}_{\phantom{\mu}\nu}$, i.e.
$R^i_{\phantom{i}i}=\sum_{i=1}^3R^i_{\phantom{i}i}$, $k_0$ is the
frequency of the  emitted photon $\gamma$, and $\eta^{00}$ is the
00-component of the metric tensor in the inertial frame,
$\eta_{\mu\nu}=(-1, 1,1,1)$. The crucial point, in order that the
\v{C}erenkov radiation be kinematically allowed, is that
$R^i_{\phantom{i}i}< 0$, so that $n^2_{\gamma}(k_0)> 1$.

In the paper \cite{gupta}, the scattering process $f(p)\to
f(p')+\gamma(k)$, responsible for the \v{C}erenkov radiation, is
analyzed in the local inertial frame of the incoming fermion
$f(p)$ with momentum $p$, while $f(p')$ and $\gamma(k)$ are the
outgoing fermion with momentum $p'$ and the emitted photon with
momentum $k$.  As one can immediately realize, \v{C}erenkov
emission is not vanishing in the inertial frame, unlike that
produced by curvature radiation, since the refractive index turns
out to be proportional to the spatial parts of the Ricci tensor.

The energy radiated by \v{C}erenkov process by a charged particle
moving in a background gravitational field is given by (for
details, see \cite{gupta})
\begin{equation}\label{1.2}
  \frac{dE}{dt}=\frac{Q^2\alpha_{em}}{4\pi p_0^2}
  \int_{k_{01}}^{k_{02}}dk_0\left[p_0(p_0-k_0)-\frac{1}{2}k_0^2\right]
  k_0\frac{n_{\gamma}^2-1}{n_{\gamma}^2}\,,
\end{equation}
where $Q$ is the charge of the fermion emitting the photon,
$\alpha_{em}$ is the electromagnetic coupling constant, $p_0$ is
the energy of the fermion, and $k_{01},k_{02}$ stands for the
allowed range of frequencies where radiation can occur.

The interesting result is obtained for $n_{\gamma}^2 \gg 1$, since
the spectrum of energy radiated by a charged particle, Eq.
(\ref{1.2}), assumes the form
\begin{equation}\label{3.10}
  \frac{d}{dk_0}\left(\frac{dE}{dt}\right)=\frac{Q^2\alpha_{em}}{4\pi}
  \left[1-\frac{k_0}{p_0}-\frac{k_0^2}{2p_0^2}\right]k_0\,,
\end{equation}
which differs in a substantial way from thermal or synchroton
emission. We show this spectrum for a monochromatic proton energy $p_0=10$
GeV in Fig. 1.
If this result
holds for boson stars, it strongly
suggest the possibility to reveal their  presence via their output in
\v{C}erenkov radiation, in addition to gravitational lensing
effects \cite{dabrowski}. Note that in most astrophysical situations, the
dense media that could give rise to \v{C}erenkov radiation are also
optically thick to the emitted radiation, making it not easily observed.
However, a transparent -uncharged- boson star offer a better 
environment for such a detection since absortion within the star will be
inexistent.

However, our main result is proof of just the opposite: no
\v{C}erenkov radiation can be produced by particles traversing a
boson star. In Sect. II we shortly review the theory underlying
boson and soliton stars, recalling the main ideas and equations for
evaluating
the refractive index, which will be the argument of Sect. III.
Conclusions are drawn in Sect. IV.

\section{Boson stars physics--Brief review}

We shall now introduce the formalism which gives rise to
mini-boson,  boson, and soliton stars. To do this we study the
Lagrangian density of a massive complex self-gravitating scalar
field, which is (taking $\hbar=c=1$)

\be\label{lagr}
 {\cal L} = \frac {1}{2} \sqrt{\mid g \mid} \left[
  \frac {m_{{\rm Pl}}^2}{8\pi} R + \partial_\mu \psi^\ast \partial^\mu \psi
 - U(|\psi |^2) \right ] \; ,
 \ee
where $R$ is the scalar of curvature, $g$ the determinant of the
metric $g_{\mu \nu }$, and $\psi$ is a {\em complex} scalar field
with potential $U$. Using this Lagrangian as the generator of the
matter sector of the theory, we get the standard field equations
 \ben
 R_{\mu \nu } - \frac{1}{2} g_{\mu \nu } R & = &
                  - \frac{8\pi}{m_{{\rm Pl}}^2} T_{\mu \nu } (\psi ) \; ,\label{2.2} \\
 \Box \psi + \frac{dU}{d|\psi |^2} \psi & = & 0 \label{2.3} \; ,
 \een
where the stress energy tensor is given by,
\begin{equation}
 T_{\mu \nu }  =  (\partial_\mu \psi^\ast ) (\partial_\nu \psi )
  - \frac{1}{2} g_{\mu \nu }
 \Bigl [ g^{\alpha \beta } (\partial_\alpha \psi^\ast )
         (\partial_\beta \psi ) - U(|\psi |^2) \Bigr ],\label{2.4}
\end{equation}
and $\Box = \partial_\mu \Bigl [ \sqrt{\mid g \mid } \; g^{\mu\nu}
\partial_\nu \Bigr ]/ \sqrt{\mid g \mid }$ is the covariant d'Alembertian.
Because of the fact that the potential is a function of the square
of the modulus of the field, we obtain a global $U(1)$ symmetry.
This symmetry is related with the conserved number of particles.
The particular form of the potential, however, is what makes the
difference between mini-boson, boson, and soliton stars.
Conventionally, when the potential is given by
 \be\label{2.5}
 U(|\psi |^2) = m^2 |\psi |^2 + \frac{\lambda }{2} |\psi |^4 \; ,
 \ee
where $m$ is the scalar mass and $\lambda$ a dimensionless
constant measuring the self-interaction strength, mini-boson stars
are those spherically symmetric equilibrium configurations with
$\lambda =0 $. Boson stars, on the contrary, have a non-null value
of $\lambda$. The previous potential with $\lambda \neq 0$ was
introduced by Colpi et al. \cite{COLPI}, who numerically found
that the masses and radius of the configurations were deeply
enlarged in comparison to the mini-boson case, even in the case of
extremely small $\lambda$.

Soliton (also called non-topological soliton) stars are different
in the sense that, apart from the requirement that the Lagrangian
must be invariant under a global $U(1)$ transformation, it is
required that --in the absence of gravity-- the theory must have
non-topological solutions; i.e.~solutions with a finite mass,
confined to a finite region of space, and non-dispersive. An
example of these kind of potentials is the one introduced by Lee
and his coworkers in a serie of 1987 papers \cite{LEE}.

We shall now briefly explain how these boson configurations can be
obtained (see Refs. \cite{rev} for details).  We adopt a
spherically symmetric line element
 \be
 ds^2 = e^{\nu (r)} dt^2 - e^{\mu (r)} dr^2
  - r^2 ( d\vartheta^2 + \sin^2\vartheta \, d\varphi^2) \; ,
 \ee
 with a scalar field time dependence ansatz consistent with
this metric:
 \be
 \psi (r,t) = \sigma(r) e^{-i \omega t} \;
 \ee
where $\omega$ is the (eigen-)frequency. This form of the field
-when the scalar has no nodes-
ensures us to be working in the configurations of minimal energy
(see appendix of Ref. \cite{LEE}).

The non-vanishing components of the energy-momentum tensor are

\ben T_0{}^0 = \rho = \frac{1}{2} [ \omega^2  \sigma^2(r) e^{-\nu
}
   + \sigma'^2(r) e^{-\mu } + U ] \; , \\
T_1{}^1 = p_r = \frac{1}{2} [ \omega^2  \sigma^2(r) e^{-\nu }
   + \sigma'^2(r) e^{-\mu } - U ] \; , \\
T_2{}^2  =   T_3{}^3  =p_\bot =   - \frac{1}{2} [ \omega^2
\sigma^2(r) e^{-\nu }
   - \sigma'^2(r) e^{-\mu } - U ],
\een where $'=d/dr$. One interesting characteristic of this system
is that the pressure is anisotropic; thus, there are two equations
of state $p_{r} = \rho - U$ and $p_\bot = \rho - U - \sigma'^2(r)
e^{-\mu }$. The non-vanishing independent components of the
Einstein equation are

\ben \nu' + \mu' & = & \frac{8\pi}{m_{{\rm Pl}}^2} (\rho + p_r) r
e^\mu \; , \label{nula}\\ \mu' & = &  \frac{8\pi}{m_{{\rm Pl}}^2}
\rho r e^\mu - \frac {1}{r} (e^\mu - 1) \; . \een Finally, the
scalar field equation is

\be
\sigma'' + \left ( \frac {\nu' - \mu'}{2} + \frac {2}{r} \right )
 \sigma' + e^{\mu - \nu } \omega^2 \sigma
- e^{\mu } \frac{dU}{d\sigma^2} \sigma = 0  \; . \ee

To do numerical computations and order of magnitude estimations,
it is useful to have a new set of dimensionless variables. The
usual ones are: $x=mr$ for the radial distance, $\Omega= \omega /
m$ for the eigenvalue, we redefine the radial part of the boson
field as $\sigma = \sqrt{4\pi} \; \sigma/m_{{\rm Pl}}$, and
introduce $\Lambda = \lambda m_{{\rm Pl}}^2/4\pi m^2$. In order to
obtain solutions which are regular at the origin, we must impose
the following boundary conditions $\sigma '(0)=0$ and $\mu (0)=0
$. These solutions have two fundamental parameters: the
self-interaction and the central density (represented by the value
of the scalar field at the centre of the star). The mass of the
scalar field fixes the scale of the problem. Boundary conditions
representing asymptotic flatness must be applied upon the metric
potentials, these determine -which is actually accomplished via a
numerical shooting method- the initial value of $\nu=\nu (0)$.
Then, having defined the value of the self interaction, or
alternatively, the form of the soliton potential, the equilibrium
configurations are parameterized by the central value of the boson
field. As this central value increases, so does the mass and
radius of the the star. This happens until a maximum value is
reached in which the star looses its stability and disperses away
(the binding energy being positive). Up to this value of
$\sigma(0)$, catastrophe theory can be used to show that these
equilibrium configurations are stable \cite{STAB}. The numerical
code used in this paper is a modification of the scalar-tensor
boson star program developed in Ref. \cite{ULT}.

\section{\v{C}erenkov radiation and scalar stars}

In this Section we calculate if \v{C}erenkov radiation (in the
sense of Gupta et al.'s paper) can be emitted by a charged
particle which feels the gravitational field generated by a boson
star. As already explained in the Introduction, in order to allow
for the emission of radiation by \v{C}erenkov process, the sum on
the spatial components of the Ricci tensor has to be negative. To
calculate it, let us rewrite the Einstein equation (\ref{2.2}) in
the following form
\begin{equation}\label{3.1}
  R^{\mu}_{\phantom{\mu}\nu}=-\frac{8\pi}{m_{Pl}^2}\left(
 T^{\mu}_{\phantom{\mu}\nu}-\frac{1}{2}\delta^{\mu}_{\nu}T\right),
\end{equation}
where $T$ is the trace of the stress--energy tensor (\ref{2.4}).
The scalar curvature is then
\begin{equation}\label{3.2}
  R=\frac{8\pi}{m_{Pl}^2}\,T=\frac{8\pi}{m_{Pl}^2}[-
  \vert\partial_{\mu}\psi\vert^2+2U(\vert\psi\vert^2)]\,
\end{equation}
where the potential $U$ is defined in (\ref{2.5}).  From
(\ref{3.1}) one can calculate the component $R^0_{\phantom{0}0}$,
getting
\begin{equation}\label{3.3}
R^0_{\phantom{0}0}=\frac{8\pi}{m_{Pl}^2}[-
  \vert\partial_0\psi\vert^2+\frac 12 U(\vert\psi\vert^2)]\,,
\end{equation}
so that the sum of the spatial components of the Ricci tensor are
given by
\begin{equation}\label{3.4}
R^i_{\phantom{i}i}=R-R^0_{\phantom{0}0}=\frac{8\pi}{m_{Pl}^2}[-
  \vert\partial_i\psi\vert^2+ \frac 32 U(\vert\psi\vert^2)]\,.
\end{equation}
Inserting Eq. (\ref{3.4}) into (\ref{1.1}), one infers the
effective refractive index
\begin{equation}\label{3.5}
  n^2_{\gamma}(k_0)=|\eta^{00}|\left[1-\frac{8\pi}{|\eta^{00}|m_{Pl}^2k_0^2}
  (-\vert\partial_i\psi\vert^2+  \frac 32 U(\vert\psi\vert^2))\right]\,.
\end{equation}
If $R^i_{\phantom{i}i}$ in Eq. (\ref{3.4}) is negative, then the
emission of \v{C}erenkov radiation is a kinematically allowed
process. The condition $R^i_{\phantom{i}i}<0$ (so that
$n^2_{\gamma}>1$) can be satisfied if one invokes the condition
that the spatial variation of the scalar field $\psi$ is greater
than three half its potential energy, i.e.
\begin{equation}\label{3.6}
 \vert\partial_i\psi\vert^2> \frac 32 U(\vert\psi\vert^2)\,.
\end{equation}

\subsection{Boson stars: Colpi et al. potential}

In order to evaluate the order of magnitude of this requirement
for boson stars let us turn into dimensionless variables the
expression: \be \frac{8\pi}{m_{Pl}^2k_0^2}
  \left[-\vert\partial_i\psi\vert^2+  \frac 32
  U(\vert\psi\vert^2)\right].\label{11}\ee
Using the adopted ansatz for the field and the previously quoted
dimensionless variables we obtain:
 \be \frac{2m^2 }{k_0^2} f(x)=\frac{2 m^2 }{k_0^2}
  \left[- \left(\frac {d\sigma}{dx}\right) ^2+  \frac 32
\left(\sigma(x)^2 + \frac{\Lambda}{2} \sigma(x)^4 \right) \right]
.\ee Thus, $n_\gamma ^2 =1- (m^2 /2 k_0^2) f(x)$. If, and when,
$f(x)$ is negative, \v{C}erenkov radiation might happen. However,
in Fig. 2 we show several numerical integrations for different
boson stars models, i.e. with different central densities and
self-interaction. It is clear that the square of the derivative of
$\sigma$ is never bigger than the terms involving the square and
the fourth power of $\sigma$ itself, and then that there is no
possible radiation.

Starting form the dimensionless expression of $f(x)$, and as in
principle $\Lambda$ could assume any value, even negative ones, we
can think of doing the following trick. Let us take those shells
very near the center of the star. There, because of the boundary
condition, $\sigma^\prime \sim 0$, and $\sigma$ itself adopts its
maximum value. Then, if $\Lambda < 0$, $f(x)$ will be negative for
all values which fulfill $|\Lambda| > 2 / \sigma(0)^2$. But
interestingly enough, this inequality can not be accomplished by
stable stars. For instance, if $\sigma(0) = 0.19$, $|\Lambda|>
55$. But as can be confirmed by Table I of Ref. \cite{LS}, for
values of $\Lambda < -20$, central densities above
$\sigma(0)=0.067$ already represent unstable solutions. Because
this happens for all values of $\sigma(0)$, this is why no
physical (stable) boson star, under Colpi et al.'s usual potential can
generate \v{C}erenkov radiation.

\subsection{Soliton stars}

It is clear that the requirement expressed by Eq. (\ref{3.6})
depends on the form of the potential. As an example of this we
study here the Lee et al.'s soliton star potential given by \be
U=m^2 |\psi|^2 \left( 1- \frac{|\psi|^2}{\chi_0^2} \right)^2,\ee
where $\chi_0$ is a constant. As we already mentioned, compared
with the usual boson star case, non-topological soliton stars have
to fulfill two characteristics: 1. The Lagrangian must be
invariant under a global $U(1)$ transformation. 2. In the absence
of gravity, the theory must have non-topological solutions. In
general, boson stars accomplish the requirement 1.~but not 2.
Invariance under $U(1)$ only requires that the potential be a
function of $\psi^* \psi$, but in order to accomplish condition
2., $U$ must contain attractive terms. This is why the coefficient
of $(\psi^* \psi)^2$ of Lee's potential has a negative sign.
Finally, when $|\psi| \rightarrow \infty$, $U$ must be positive,
which leads, minimally, to a sixth order function of $\psi$ for
the self-interaction.

As we did above, we shall obtain the dimensionless form for
expression (\ref{11}), but in this case for the soliton potential.
It is given by (without multiplying factors)  \be g(x)=-
\left(\frac {d\sigma}{dx}\right) ^2+ \frac 32 \left(\sigma(x)^2 -2
\alpha^2 \sigma(x)^4 + \alpha^4 \sigma^6 \right),\ee where $\alpha
= m_{Pl}^2/ 4 \pi \chi_0^2$. Note that the parameter $\alpha$ does not
directly depend on the mass of the boson, but just on the 
particular value of the $\chi_0$ parameter. It is usually
assumed,
however, that this parameter is of the order of the boson mass.

We can now think of the following
situation: as $\sigma$ depends on $x$, for values of $\sigma=1/\alpha$,
the second term is
exactly zero, and then $g(x)$ will be negative if the derivative
of the field is not null. This can be easily obtained starting
from a central value of $\sigma$ slightly higher than $1/\alpha$
(so as to pass through $\sigma=1/\alpha$ well inside the structure
of the star, where the derivative is not zero). In Fig. 3 we show
different models fulfilling these constraints. We may see that the
higher the value of $\alpha$, the more dilute the center of the
star is. It is also possible to see that for bigger values of
$\alpha$, $|g(x)|$ takes even smaller values. Note too that a
particular choice for the parameters (basically, for the central
density) determine the spherical shell in which $g(x)$ is
negative, and thus where \v{C}erenkov radiation might occur. For
instance, if $\alpha$ is very big, but $\sigma(0)$ is also very
large, we could still find \v{C}erenkov radiation in the outer
``crust'' of the star.

This shows an, in principle, interesting observable difference
between boson and non-topological stars, which is caused by the
form of the potential. The actual prospects for really observing
this are to be determined particularly by the value of the boson
mass involved. If $m$ is about 30 GeV, and $\chi_0$ is of the
order of the boson mass, the parameter
$\alpha$ is quite large, and the observational possibility diminishes,
since 
\v{C}erenkov radiation 
would happen only for extremely dilute stars (the concept of star
itself looses sense in this situation). We have checked, however,
that the negativeness of $g(x)$ is numerically preserved up to
values of $\alpha=10^4$. But is so small as $10^{-10}$. Observing
radiation generated in these conditions would be quite a different
story. However, we may note that there could be a sort of fine tunning
in the values of $\chi_0$ and $m$ for which $n_\gamma^2=1-(m^2/2k_0) 
g(x)$ be large. If this fine tunning has any physical
motivation should be decided on particle physics grounds. Additionally, we
mention that for values of $m_{Pl}/m \sim 10^{17}$, it
is needed a different numerical technique: basically we need to solve the
equations in three separate parts, making adequate expansions (see
the papers by Lee et al.'s for details \cite{LEE}). In those
cases, the mass and radius of the stars are enlarged up to galaxy
and galaxy cluster masses within some light years of linear
spread. We have not taken these cases here into account.

\section{Final remarks}

In this paper we have analyzed the possibility that a charged
particle, moving in a gravitational field generated by a scalar
star, could emit radiation through the gravitational \v{C}erenkov
process recently introduced by Gupta et al. in Ref. \cite{gupta}.
We have numerically shown that the usual boson star model, based
on the potential introduced in this context by Colpi et al., is
not able to generate a refractive index bigger than 1, and thus,
that \v{C}erenkov radiation can not proceed. We have also shown
that, on the contrary, soliton stars could have \v{C}erenkov
radiation for particular values of the boson mass. However,
diluteness of the star grows and actual observational possibility
decreases for the more usually assumed boson masses and $\chi_0$
parameters. For these cases, $(n_\gamma^2 -1) \sim 0$ and \v{C}erenkov
process might be considered as experimentally insignificant within current
observational constraints, on a par with what Gupta et al. 
have concluded for the galactic gravitational field \cite{gupta}. We
may think of a possibility to 
overcome the smallness of $(n_\gamma^2 -1)$ if a soliton
star is alligned with a strong proton source. A more, maybe unexpected
form, would be to find soliton stars with the right combination of
$\chi_0$ and $m$, as to have a large effective refractive index. Overall,
it appears that 
other ways 
-apart from 
the already mentioned gravitational lensing effects \cite{dabrowski}- 
have to be devised in order to look for scalar stars.

\subsection*{Acknowledgments}

We thank Drs. L. A. Anchordoqui, F. E. Schunck, S. Mohanty, as well as two
anonymous referess, for criticism
and valuable suggestions. S. C. and G. L.'s research was supported by
MURST fund (40\%) and
art. 65 D.P.R. 382/80 (60\%). G.L. further thanks UE (P.O.M.
1994/1999). D.F.T. was supported by CONICET as well as by funds
provided by Fundaci\'on Antorchas, and acknowledges the hospitality
provided by the ICTP (Italy) during the latest stages of this work.


\newpage

\begin{figure}
 \leavevmode \epsfxsize=8.5cm \epsfysize=9.5cm
\epsffile{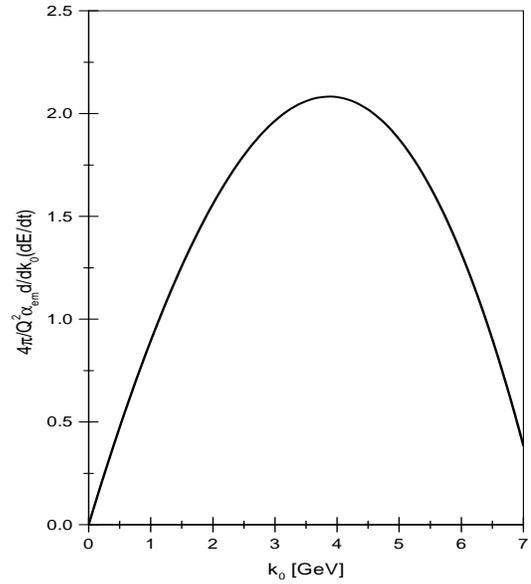}
 \caption{Typical spectrum for the radiation emitted by monochromatic 
charged
particles  
with $|p|= 10$ GeV in a gravitational field with effective refractive
index bigger than 1, as was first obtained by Gupta et al. in Ref. [19].} 
\label{Fig1}
\end{figure}

\begin{figure}
 \leavevmode \epsfxsize=8.5cm \epsfysize=9.5cm
\epsffile{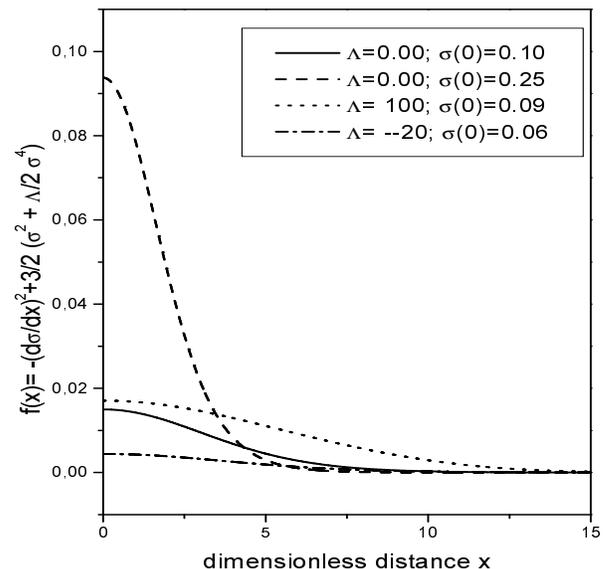}
 \caption{Numerical integration of the boson star structure equations
 for different central densities and self-interaction.
 All of them represent stable configurations.
 As $f(x)$ is always positive, \v{C}erenkov radiation can not be
 present.}
\label{Fig2}
\end{figure}

\begin{figure}[t]
 \leavevmode \epsfxsize=8.5cm \epsfysize=10.5cm
\epsffile{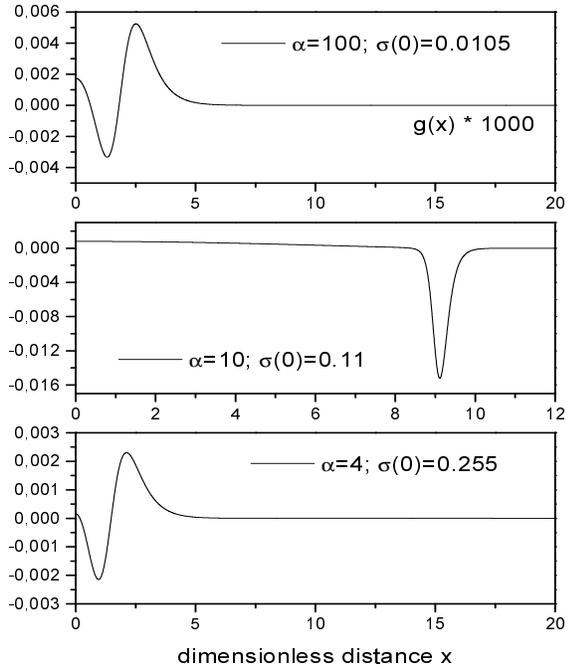}
 \caption{Numerical integration of soliton star structure equations
 for different central densities and self-interaction parameters.
 We show here the function $g(x)$ involved in the computation of the
 refractive index: where $g(x)$ is negative, \v{C}erenkov
 radiation might occur. The difference in the form of the function
 between the middle and the bottom and top plots occurs when the departure of
 $\sigma(0)$ from the critical value $1/\alpha$ is bigger.}
\label{Fig3}
\end{figure}

\end{document}